# Citation Counting, Citation Ranking, and *h*-Index of Human-Computer Interaction Researchers: A Comparison between Scopus and Web of Science

**Lokman I. Meho**
School of Library and Information Science, Indiana University Bloomington
Bloomington, IN 47405; meho@indiana.edu

**Yvonne Rogers**
Department of Computing, Open University, United Kingdom; Y.Rogers@open.ac.uk

**Abstract:** This study examines the differences between Scopus and Web of Science in the citation counting, citation ranking, and *h*-index of 22 top human-computer interaction (HCI) researchers from EQUATOR—a large British Interdisciplinary Research Collaboration project. Results indicate that Scopus provides significantly more coverage of HCI literature than Web of Science, primarily due to coverage of relevant ACM and IEEE peer-reviewed conference proceedings. No significant differences exist between the two databases if citations in journals only are compared. Although broader coverage of the literature does not significantly alter the relative citation ranking of individual researchers, Scopus helps distinguish between the researchers in a more nuanced fashion than Web of Science in both citation counting and *h*-index. Scopus also generates significantly different maps of citation networks of individual scholars than those generated by Web of Science. The study also presents a comparison of *h*-index scores based on Google Scholar with those based on the union of Scopus and Web of Science. The study concludes that Scopus can be used as a sole data source for citation-based research and evaluation in HCI, especially if citations in conference proceedings are sought and that *h* scores should be manually calculated instead of relying on system calculations.

## INTRODUCTION

Citation analysis—i.e., the analysis of data derived from references cited in footnotes or bibliographies of scholarly publications—is a powerful and popular method of examining and mapping the intellectual impact of scientists, projects, journals, disciplines, and nations (Borgman, 1990; Garfield, 1979; Meho, 2007; Moed, 2005). The method is increasingly being used by academic, research, and federal institutions in several countries worldwide for research policy making, visualization of scholarly networks, and monitoring of scientific developments, as well as for promotion, tenure, hiring, salary raise, and grants decisions, among others (see Borgman & Furner, 2002; Cronin, 1996; Small, 1999; Warner, 2000; Weingart, 2005; White & McCain, 1997, 1998). Indeed, several governments have been using or are considering using citation analysis and other bibliometrics measures/indicators to inform decisions regarding research quality assessment and the allocation of research funds in higher education (see Adam, 2002; Butler, 2007; Moed, 2007, 2008; Weingart, 2005).

Major reasons for the growing popularity of citation analysis include: (1) the validity and reliability of the method in assessing, supporting, or questioning peer review-based judgments regarding the impact of a scientist's research output, especially in domains where the journal article and conference paper are considered the main scholarly communication channels; (2) the relative ease with which one can collect citation data; (3) the proliferation of several bibliometrics products (e.g., ISI Essential Science Indicators—http://www.in-cites.com/rsg/—and ISIHighlyCited.com), tools (e.g., Scopus and Google Scholar), and measures (e.g., *h*-index and *g* index) which can facilitate citation-based research and evaluation; (4) the ability of the method to create competition among academic and research institutions (by way of rankings) and thus increase their efficiency; and (5) the growing skepticism and disenchantment with peer review as a sole research evaluation method (for more on last point, see Norris & Oppenheim, 2003; Rinia, van Leeuwen, van Vuren, & van Raan, 1998; Weingart, 2005).

The basic idea or assumption behind citation analysis is that influential works or scientists are cited more often than others. In this sense, citations reflect the relative impact and utility of a work, author, department, or journal's publications within their larger scientific domains. Because the quality, validity and reliability of citation-based research and evaluation is highly dependent on the accuracy and comprehensiveness of the data used, it is imperative that appropriate citation sources and data collection methods are utilized (see van Raan, 1996, 2005; Weingart, 2005). Inaccurate or incomplete data risks underestimating the impact of a scientist, department, university, journal, or nation's research output that may otherwise be deemed good by established standards.

Until recently, most citation-based research relied exclusively on data obtained from Web of Science, which consists of three Institute for Scientific Information (currently Thomson Scientific) citation databases: Arts & Humanities Citation Index, Science Citation Index, and Social Sciences Citation Index. The emergence Elsevier's Scopus database in late 2004, however, has raised many questions regarding: (1) the validity of findings based exclusively on data from Web of Science; (2) the value and necessity of using multiple citation sources for examining and mapping the intellectual impact of research; and (3) the appropriateness of using Scopus as an alternative source of citations to Web of Science. These three issues are raised primarily because of the considerably broader literature coverage in Scopus (over 15,000 "peer-reviewed" titles, including more than 1,000 Open Access journals, 500 conference proceedings, and 600 trade publications going back to 1996) than that of Web of Science (approximately 9,000 scholarly journals and a significant number of conference proceedings and books in series); users of citations for research evaluation want to know what are the effects of this broader coverage on evaluation results, how significant are the effects of this broader coverage, and what characterizes the sources exclusively covered by Scopus (in terms of impact, quality, and type of documents).

**LITERATURE REVIEW**

Studies that explored the differences between citation sources had different results. For example, Bauer and Bakkalbasi (2005) compared citation counts provided by Scopus, Google Scholar, and Web of Science for articles from the *Journal of the American Society for Information Science and Technology* published in 1985 and in 2000. The results for 1985 articles were inconclusive, but for 2000 articles, Google Scholar provided statistically significant higher citation counts than either Scopus or Web of Science. The authors concluded that researchers should consult Google Scholar in addition to Scopus or Web of Science, especially for relatively recent publications, but until Google Scholar provides a complete accounting of the material that it indexes and how often that index is updated, Google Scholar cannot be considered a true scholarly resource in the sense that Scopus and Web of Science are.

Jacsó (2005) conducted several tests comparing Google Scholar, Scopus, and Web of Science, searching for documents citing (a) Eugene Garfield, (b) an article by Garfield published in 1955 in *Science*, (c) the journal *Current Science*, and (d) the 30 most-cited articles from *Current Science*. He found that coverage of *Current Science* by Google Scholar is "abysmal" and that there is considerable overlap between Scopus and Web of Science. He also found many unique documents in each source, pointing out that the majority of the unique items were relevant and substantial. Noruzi (2005) studied the citation counts in Google Scholar and Web of Science of 36 webometrics papers; in most cases, he found that Google Scholar provided higher citation counts than Web of Science. These findings were corroborated by the results of Vaughan and Shaw (2008) for information science.

Bakkalbasi, Bauer, Glover, and Wang (2006) compared citation counts for articles from 11 oncology journals and 11 condensed matter physics journals published in 1993 and 2003. They found that for oncology in 1993, Web of Science returned the highest average number of citations (45.3), Scopus returned the highest average number of citations for oncology in 2003 (8.9), and Web of Science returned the highest number of citations for condensed matter physics in 1993 and 2003 (22.5 and 3.9, respectively). Their data showed a significant difference in the mean citation rates between all pairs of resources except between Google Scholar and Scopus for condensed matter physics in 2003. For articles published in 2003, Web of Science returned the largest amount of unique citing material for condensed matter physics and Google Scholar returned the most for oncology. The authors concluded that all three tools returned some unique material and that the question of which tool provided the most complete set of citing literature might depend on the subject and publication year of a given article. In four science disciplines, Kousha and Thelwall (2006) found that the overlap of citing documents between Google Scholar and Web of Science varies from one field to another and, in some cases, such as chemistry, it is relatively low (33%).

Norris and Oppenheim (2007) used all but 720 of the journal articles submitted for the purpose of the 2001 Research Assessment Exercise in the social sciences (n=33,533), as well as the list of 2,800 journals indexed in the *International Bibliography of the Social Sciences*, to assess the coverage of four data sources (CSA Illumina, Google Scholar, Scopus, and Web of Science). They found that Scopus provides the best coverage of social science literature from among these data sources and concluded that Scopus could be used as an alternative to Web of Science as a tool to evaluate research impact in the social sciences. Bar-Ilan (2006) carried out an ego-centric citation and reference analysis of the works of the mathematician and computer scientist, Michael O. Rabin, utilizing and comparing Citeseer, Google

Scholar, and Web of Science. She found that the different collection and indexing policies of the different data sources lead to considerably different results. In another study, Bar-Ilan, Levene, and Lin (2007) compared the rankings of the publications of 22 highly-cited Israeli researchers as measured by the citation counts in Google Scholar, Scopus, and Web of Science. The results showed high similarity between Scopus and Web of Science and lower similarities between Google Scholar and the other databases. More recently, Bar-Ilan (2008) compared the *h* scores (see below) of a list of 40 highly-cited Israeli researchers based on citation counts from Google Scholar, Scopus, and Web of Science. In several cases, she found that the results obtained through Google Scholar were considerably different from those in Scopus and Web of Science, primarily due to citations covered in non-journal items.

Meho and Yang (2007) used citations to more than 1,400 works by 25 library and information science faculty to examine the effects of additionally using Scopus and Google Scholar on the citation counts and rankings of these faculty as measured by Web of Science. The study found that the addition of Scopus citations to those of Web of Science significantly altered the relative ranking of those faculty in the middle of the rankings. The study also found that Google Scholar stands out in its coverage of conference proceedings as well as international, non-English language journals. According to the authors, the use of Scopus and Google Scholar, in addition to Web of Science, reveals a more comprehensive and complete picture of the extent of the scholarly relationship between library and information science and other fields.

In addition to the above studies, there are several papers that focused on the variations in coverage, user friendliness, and other advantages and disadvantages of Google Scholar, Scopus, and/or Web of Science, most recently: Falagas, et al (2008), Golderman and Connolly (2007), and Goodman and Deis (2007). These papers and the studies reviewed suggest that the question of whether to use Scopus and/or Web of Science as part of a research assessment exercise might be domain-dependent and that more in-depth studies are needed to verify the strengths and limitations of each data source.

**RESEARCH PROBLEM**

Building on previous research, this study examines the differences in coverage between Scopus and Web of Science for the particular domain of human-computer interaction (HCI). HCI, which intersects both the human and computer sciences, is concerned with "designing interactive products to support the way people communicate and interact in their everyday and working lives" (Sharp, Rogers, & Preece, 2007, p. 8) and "with the study of major phenomena surrounding them" (Hewett et al, 1992, p. 5). It should be emphasized here that HCI is synonymous with CHI (computer–human interaction), a term or acronym that was essentially used in the U.S. Researchers and practitioners more generally and internationally now refer to the domain as HCI (see Grudin, 2008). According to Dillon (1995) and Valero and Monk (1998), HCI emerged from a supporting base of several disciplines, including, computer science, information systems, cognitive and organizational psychology, and human factors. Shneiderman and Lewis (1993) indicated major influences by business, education, and library and information science departments too. Given this broad base and the diversity of places where HCI researchers publish, it could be that there are marked differences in coverage of HCI citation literature between Scopus and Web of Science. To investigate if this is the case, we look at the differences between the two databases for the citation counting, citation ranking, and $h$-index scores of 22 top HCI researchers from a large British Interdisciplinary Research Collaboration project, called EQUATOR. More specifically, the study addresses three questions:

- How do the two databases compare in their coverage of HCI literature and the literature that cites it, and what are the reasons for the differences?

- What impact do the differences in coverage between the two databases have on the citation counting, citation ranking, and $h$-index scores of individual HCI researchers?

- Should one or both databases be used for determining the citation counting, citations ranking, and $h$-index scores of HCI researchers?

The $h$-index, a relatively new bibliometric measure, was developed by physicist Jorge Hirsch (2005) to quantify the impact of individual scientist's research output and correct for various perceived deficiencies of citation counting and ranking methods. Unlike citation counting and ranking, which can be easily influenced by one or very few highly cited papers or by the number of papers a scientist has published regardless of their quality, the *h*-index takes into account both the quantity and "quality" (or impact) of publications and helps to identify distinguished scientists who publish a considerable number of highly cited papers. The formula for the *h*-index is simple: A scientist has an index *h* if *h* of his or her papers have at least *h* citations each. That is to say, a scientist with an *h*-index of 10 has published 10 works that have each attracted at least 10 citations. Papers with fewer than 10 citations don't count. Like any other citation-based measure, the *h*-index has several weaknesses, perhaps most importantly is the fact that it does not take into account the total number of citations an author has accumulated. It also cannot be used to make cross-disciplinary comparisons. For example, many physicists can and have achieved an *h* score of 50 or more (Hirsch, 2005), whereas in such fields as library and information science (LIS) very few have reached the score of 15 based on data from Web of Science (Cronin & Meho, 2006; Oppenheim, 2007). For more on the *h*-index and the various models used to improve it, see Bar-Ilan (2008a), Bornmann, Mutz, and Daniel (2008), and Jin, Liang, and Rousseau (2007).

Unlike previous *h*-index studies, which exclusively relied on *h* scores computed by the database system, the current study calculates, compares, and uses two types of *h* scores: *system count* and *manual count*. In the system-based counting method, *h* scores are determined by identifying all papers indexed in a database for an author and then using the "Citation tracker" and "Citation Report" analytical tools in Scopus and Web of Science, respectively, to calculate the *h* scores. In this method, the *h* scores will not take into account an author's cited works that are not covered by the database. In contrast to the system-based *h*-index count, in the manually-based counting method, *h* scores are calculated by identifying the citation count of each work by an author regardless of whether the work is indexed in a database. This is followed by ranking the works by most cited first, then counting down until the number of times cited equals or is less by one than the number of cited works. To our knowledge, very few studies have compared these two types of counting methods (e.g., Cronin & Meho, 2006). Similarly, very few studies have compared Scopus and Web of Science in terms of author *h*-index (e.g., Bar-Ilan, 2008b; Sanderson, in press).

Answering the abovementioned research questions and examining the differences between system-based and manually-based *h*-index scores are important because it will allow us to more reliably rate Scopus as a data source against Web of Science. If differences are found between domains, people who use citation analysis for research evaluation and other purposes will need to justify their choice of database. Simply claiming that Web of Science is the established source will no longer be sufficient. Moreover, because citation-based metrics (e.g., citation counting or ranking, citations per paper, journal impact factors, and *h*-index) are often used in research evaluation, literature mapping, and research policy making, as well as in hiring, promotion and tenure, salary raise, and research grants decisions, it is important to determine whether citation searching in HCI and beyond should be extended to both Scopus and Web of Science or limited to one of them.

**STUDY SAMPLE**

In order to examine the differences between Scopus and Web of Science in the citation counting, citation ranking, and *h*-index scores of HCI researchers, we used a sample of 22 top scholars (11 principal investigators and 11 research fellows) from a large United Kingdom (UK) multi-institution Interdisciplinary Research Collaboration funded project known as EQUATOR (http://www.equator.ac.uk/). EQUATOR was a six-year (2001-2007) Interdisciplinary Research Collaboration (IRC), supported by the UK's Engineering and Physical Sciences Research Council (EPSRC), which focused on the integration of physical and digital interaction. It comprised a group of leading academic researchers in the design, development, and study of interactive technologies for everyday settings from eight UK universities. The expertise of the IRC was diverse, including hardware engineering, computer graphics, mobile multimedia systems, art and design, software development and system architecture, information sciences, and social and cognitive sciences. About 200 people worked on or were associated with EQUATOR; each university site had between 20-30 researchers during its lifetime, in the form of principal investigators, doctoral students, research fellows, and visiting scientists from outside of the UK.

A recent study by Oulasvirta (2007) ranked two of the study sample researchers, Benford and Gaver, among the top 20 most published and most cited authors in the Association for Computing Machinery's 1990-2006 proceedings of the Conference on Human Factors in Computing Systems (CHI), widely considered the top conference in HCI; a third researcher, Cheverst, ranked 43rd. Benford, Rodden, and Rogers have also been consistently featured among the top 100 published authors in Gary Perlman's *HCI Bibliography* of most published HCI authors (http://www.hcibib.org/authors.html). Eleven other study sample members are featured in the bibliography, too, which included in February 2008 approximately 1,500 authors with 10 or more publications in the domain.

In total, the 22 researchers included in this study had published or produced (through December 2007) 1,440 works (excluding meeting abstracts, presentations, book reviews, and 1-2 page-long editorials), which consisted of 967 (67%) conference/workshop papers; 348 (24%) journal/review articles, including cited magazine articles; 49 (3.5%) book chapters; 25 (2%) edited books and conference proceedings; 22 (1.5%) dissertations; 18 (1%) published and/or cited technical reports; and 11 (1%) books. Of these 1,440 unique items, 594 (41%) are covered by Scopus and 296 (21%) by Web of Science. Merging the results from both databases increases the number of covered items to 647 (45%). Further examination of the results shows that Scopus covers 39% of all conference papers and 61% of all journal articles published by the researchers, in comparison to Web of Science's 11% and 54%, respectively.

Although the 22 researchers were not selected randomly, it should be emphasized that when forming the EQUATOR research team, considerable attention was paid to representation by distinguished scholars who represented the primary HCI research areas, including computer science, engineering, and psychology, among others. Table 1 provides the name, the year the doctoral degree was earned, the name of the university granting the doctoral degree, and the academic/disciplinary background of the 22 researchers constituting the study sample. While we do not claim that our findings can be generalized to the whole of the HCI community, especially because American and European research focuses on information technology and people may differ in important ways (see Galliers & Whitley, 2002), we believe that our sample provides valuable information regarding the differences between Scopus and Web of Science and whether one or both databases should be used in citation-based research and evaluation in HCI.

**DATA COLLECTION**

In Scopus, we used three searching methods to determine the researchers' $h$ scores and their total citation counts: Author Search, the "More" tab, and exact match. In the first method, we identified for each individual researcher all his or her publications in the database and recorded and retrieved all the citations to these publications as automatically generated by the database. In the second method, we used the "More" searching/browsing feature to display, select, and collect citation data to items not found through or covered by the Author Search method (examples of these items are books, chapters in books, technical reports, dissertations, and journal articles and conference papers not indexed by the database). In the exact match search method, we used the title of an item as a search statement (e.g., *The Human-Computer Interaction Handbook*) and tried to locate an exact match in the cited "References" field of the indexed records. In cases where the title was too short or ambiguous to refer to the item in question, we used additional information as keywords (e.g., the first author's last name) to ensure that we retrieved only relevant citations. In cases where the title was too long, we used the first few words of the title because utilizing all the words in a long title may increase the possibility of missing some relevant citations due to typing or indexing errors. The "exact match" search method was most practical for authors with common last names (e.g., B. Brown, H. Muller, and A. Schmidt), whereas the combination of Author and "More" search methods was more practical for authors with less common last names. In Web of Science, we used the "Cited Reference Search" method to identify both citations to all 1,440 items in our sample and the researchers' $h$ scores. When necessary, we used different permutations and search strategies to ensure that we captured all relevant citations.

An important consideration in HCI, especially with regard to calculating the $h$-index, is the multiple manifestations of a work, i.e., its publication in several venues (e.g., technical reports, conference proceedings, journals, collections). In this study, we treated two different versions of works with the exact same title as one item, especially when they were produced and/or published within one year from each other; on average, there were approximately two such cases per researcher. The implications of multiple manifestations of a work for citation analysis are discussed extensively in Bar-Ilan (2006).

To carry out the study, we requested from and were provided with the complete lists of publications for our sample of 22 researchers. Although the lists seemed to be complete, we examined them with searches in several online databases/sources with extensive coverage of HCI literature (e.g., ACM Digital Library, Ei Compendex, IEEE Xplore, Inside Conferences, INSPEC, SpringerLink, Pascal, PsycINFO, Scopus, and Web of Science, as well as Google Scholar and WorldCat). This check identified 71 works that were cited (in some cases over 10 times) but were missing from the lists of publications that were provided (e.g., short conference papers, articles in professional magazines, and technical reports). The check also identified 45 citation errors (mostly in the title field, followed by author, and publication year). The use of complete and accurate publication lists helped ensure that we conducted complete citation searching and generated accurate citation counts and *h* scores. The importance and value of the use of publication lists in citation analysis is well described in Jacsó (2006) who shows that citation counts can be considerably deflated because citations to a work or an author are not grouped together automatically.

The data were collected twice—in March 2007 and again in February 2008 to ensure accuracy and currency. The citations were entered into an Excel spreadsheet and Access database and were coded by first author, source (e.g., journal and conference name), document type (e.g., journal article, review article, conference paper), reference type (e.g., journal vs. conference proceeding), publication year, language, institutional affiliation of the correspondence author, and country of the correspondence author, as well as the source used to identify the citation. Virtually all citations were from refereed sources. Approximately 3% of the citations did not have country and institutional affiliation information. We painstakingly used the Web to identify missing information. Because some journal and conference names are not entered consistently in Scopus and Web of Science (e.g., *Information Research* is indexed as *Information Research* in Scopus whereas it is *Information Research-An International Electronic Journal* in Web of Science), we manually standardized all such instances. In cases where a citing source had changed its name, we merged the citations under their most recent respective name (e.g., citations found in the *Journal of the American Society for Information Science* were listed under its more recent name, the *Journal of the American Society for Information Science and Technology*).

**RESULTS AND DISCUSSION**

The results of this study are presented and discussed in four sections: (1) the differences between Scopus and Web of Science in their coverage of the citing literature and the reasons for these differences; (2) the impact of differences in coverage of the citing literature on citation counting, citation ranking, and *h*-index scores of HCI researchers and the wisdom and value of using both databases for these purposes; (3) the differences between Google Scholar and the union of Scopus and Web of Science in terms of *h* scores and the reasons for these differences; and (4) conclusions and implications. Because Scopus and Web of Science provide different citation coverage periods, we limited the analysis to citations from years common to both databases, i.e., 1996 on—there were 255 citations from the pre-1996 period, all found in Web of Science.

**Differences in Coverage of Citing Literature**

Our results show that, in total, the 22 sample members have been cited in 7,439 different documents published between 1996 and 2007. Of these, Scopus covers 6,919 (93%) whereas Web of Science covers 4,011 (54%) (see Figure 1). A principal reason why Scopus finds significantly more citations than Web of Science is due to its coverage of significantly more citing conference proceedings: 775 in comparison to 340, respectively (see Figure 2 and, for more detail, Table 2). The impact of wider coverage of conference proceedings by Scopus on the citation results in this study is further evidenced by the considerably high number of unique citations found in conference proceedings in comparison to citations found in journals. Approximately 76% (2,596) of all citations found in conference proceedings were unique to a single database in comparison to 34% (1,352) in the case of citations in journals (see

Table 3). Similar conclusions were drawn when comparing overlap in citations in conference proceedings with those in journals (see Table 4). The prominence of conference proceedings as a major source of citations in HCI should not be surprising here, especially because of the close ties between the domain and computer science, a field that considers peer-reviewed conference proceedings as important if not more important than scholarly journals (see Bar-Ilan, 2008b; Goodrum, McCain, Lawrence, & Giles, 2001; Moed & Visser, 2007).

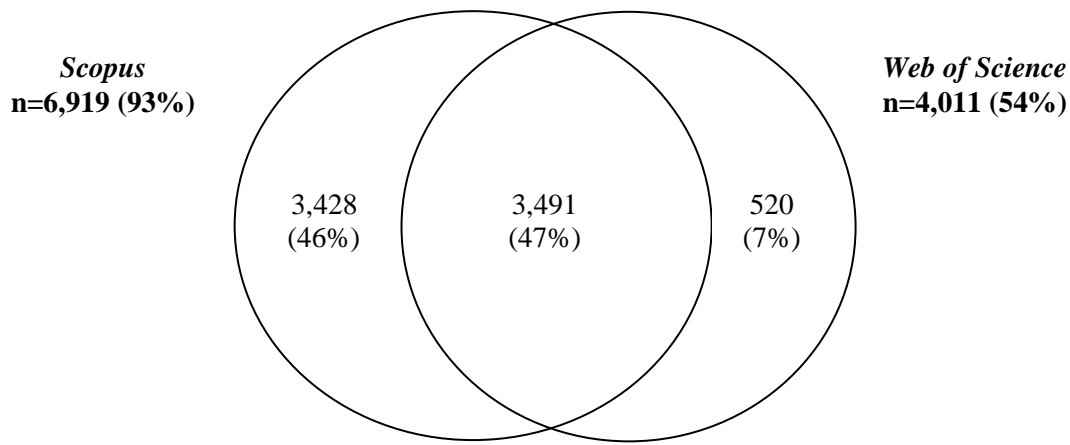

**Figure 1. Distribution of unique and overlapping citations in *Scopus* and *Web of Science* (N=7,439)**

While these findings suggest that, for HCI, more valid citation analyses are likely to be obtained through using Scopus than Web of Science, it is important to emphasize that wider coverage is not necessarily better because it may mean coverage of lower quality publications. It is often argued in academic circles that citations in high quality publications and/or from prominent authors and institutions carry more weight or are more valuable than citations found in low impact publications, and, therefore, sources of citations should be examined in order to assess the true value of the citations, especially when used in an evaluation exercise (see Neary, Mirrlees, & Tirole, 2003; Palacios-Huerta & Volij, 2004; Pinski & Narin, 1976). Given both the fact that Web of Science is the more well established citation database and the claim that it covers only or mainly high impact journals, we decided to assess the status of the sources in which Scopus's citations were found. We focused on the top 20 citing journals and 20 conference proceedings. Our assumption is that the top citing journals and conference proceedings are the most important channels of scholarly communication in a given domain and, therefore, it is expected that these journals and conference proceedings are being indexed in citation databases. This assumption is actually one of the main criteria for journal selection in Web of Science (Ball & Tunger, 2006; Testa, 2004).

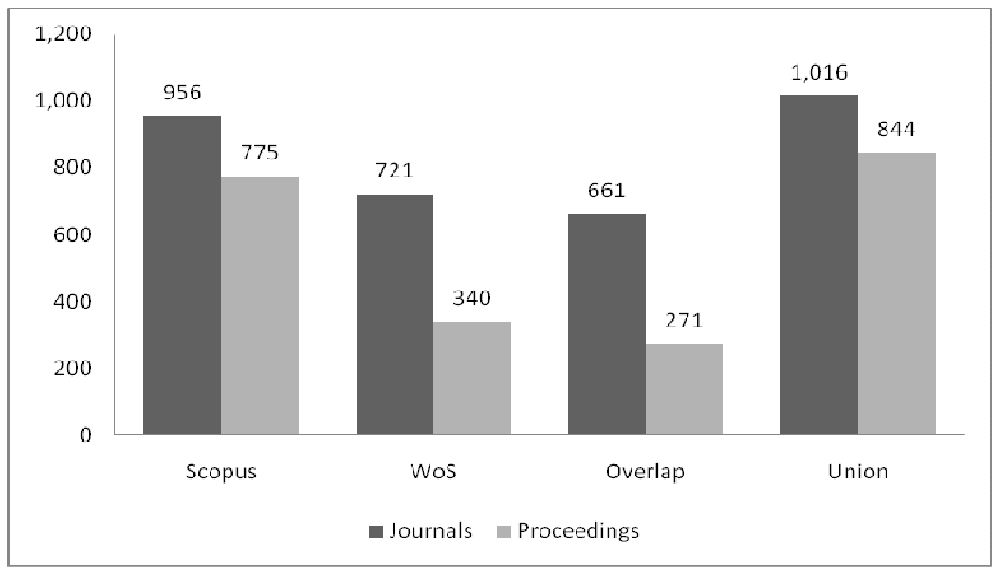

**Figure 2. Number of citing journals and conference proceedings by data source**

Our results show that Scopus covers all of the top 22 citing journals and 20 conference proceedings, in comparison to 19 journals and eight conference proceedings in the case of Web of Science; we used 22 journals instead of 20 because of a tie at rank 20 (see Table 5). These 42 journals and conference proceedings represent 2% of all citing sources and account for 30% of all citations of the study sample in both databases. Table 5 further shows that seven of the 12 conference proceedings uniquely covered by Scopus are published by ACM (the Association for Computing Machinery) and four by IEEE (Institute of Electrical and Electronics Engineers), two major sources of HCI and other fields' literatures that are widely known to publish papers of "sufficiently high level of quality" and those that are "seriously refereed" (Moed & Visser, 2007, p. vi). Table 5 also indicates that a third major source of citations in HCI is the *Lecture Notes in Computer Science/Lecture Notes in Artificial Intelligence* series, which are covered by both Scopus and Web of Science.

The impact (or "quality") of the top citing journals and conference proceedings uniquely covered by Scopus (n=15) was compared with the 27 top citing titles covered by both Scopus and Web of Science. We found that several of them have relatively high impact factor rankings/scores, including ACM Transactions on Computer-Human Interaction (3rd) and Computer Supported Cooperative Work (4th) among journals, and ACM Conference on Human Factors in Computing Systems (1st), IEEE International Conference on Pervasive Computing and Communications, PerCom (3rd), ACM Conference on Hypertext and Hypermedia (4th), and IEEE Virtual Reality Conference (5th) among conference proceedings (see Table 5 and below for other examples).

To investigate whether Web of Science covers any high impact, frequently citing journals and conference proceedings not indexed in Scopus, we analyzed the 520 citations found exclusively in Web of Science. Results showed that 322 (62%) of these citations were in sources covered by Scopus, such as Ubicomp, IEEE Pervasive Computing, ACM Computing Surveys, Interacting with Computers, Computer Networks, Journal of the American Society for Information Science and Technology, and ACM International Conference on Human-Computer Interaction with Mobile Devices and Services (MobileHCI). Scopus missed these 322 citations because of errors in the database (e.g., providing incomplete lists of cited references, lack of cited references information, and errors in cited reference information in some of its records) or because of incomplete coverage of periodicals (e.g., missing the coverage of some issues or volumes of a title or dropping the coverage of certain titles). The remaining 198 Web of Science unique citations were found in too many sources (60 journals and 69 conferences) to identify prominent and frequently citing journals and conference proceedings.

Similarly, to investigate whether Scopus covers any high impact, frequently citing journals and conference proceedings not indexed in Web of Science (and apart from those 15 Scopus unique titles that featured among the top 42 discussed earlier), we analyzed the 3,428 citations found exclusively in Scopus. Results showed that 533 (16%) of them were in sources covered by Web of Science; Web of Science missed these 533 citations primarily because of incomplete coverage of some titles. The remaining 2,895 citations found exclusively in Scopus were from 296 journals and 506 conferences—two that stood out among these 802 titles as frequently citing sources (over 20 citations each) were: the Proceedings of the Annual ACM Symposium on User Interface Software and Technology (UIST), which has a 2006 impact factor score of 2.264, and the International Conference on Intelligent User Interfaces (IUI), which has a 2006 impact factor score of 1.391. For more examples, see Table 5.

The findings presented above underline the importance of conference proceedings as a major scholarly communication channel in HCI. This was not surprising given the fact that HCI borrows from and exports ideas to several domains that rely heavily on conferences, such as computer science (see Bar-Ilan, 2006, 2008b; Goodrum, McCain, Lawrence, & Giles, 2001; Moed & Visser, 2007). The findings also show evidence that Scopus provides significantly more comprehensive coverage of HCI literature than Web of Science, primarily in terms of conference proceedings. It should be emphasized here, however, that Web of Science "intentionally" has a very poor coverage of proceedings and, had we limited our analysis to citations in "high-impact" journals only, our results would have suggested more comparable literature coverage between the two databases. Still, in order to provide better journal coverage in HCI, this study recommends that Web of Science and JCR further expand their HCI literature coverage with at least the following two prominent HCI titles: *ACM Transactions on Computer-Human Interaction* and *Computer Supported Cooperative Work*.

The effects of our findings on citation counting, citation ranking, and *h*-index scores of HCI scholars are discussed below. Because Scopus's coverage of HCI research and the literature that cites it is significantly higher than that of Web of Science, the discussion concentrates on the wisdom, necessity, and/or value of using Web of Science as an additional source of citation data. This decision was additionally driven by the fact that Scopus indexes all of the top citing publications found in Web of Science, as well as several key, high-impact HCI journals and conference proceedings that were not found in Web of Science.

**Differences in Citation Counting, Citation Ranking, and *h*-Index**

Given that Scopus covers 93% of all citations in comparison to Web of Science's 54%, it was not surprising to find that Scopus identifies significantly higher citation counts for all 22 researchers than Web of Science does, with considerable variations from one researcher to the other (ranging from a low 55% increase/difference to a high 140%). Despite this, results show that both databases produce very similar citation rankings of the 22 researchers (Spearman rank order correlation coefficient for the two rankings=0.970) (see Table 6). Results also show that the addition of citations from one database to those of the other does not significantly change the rankings. These findings suggest that the selection and use of a particular citation database will depend on the purpose of a study. If the purpose is only to compare the ranking of HCI scholars, then either database can be used, with Web of Science being the choice if citations prior to 1996, the period Scopus does not cover, are sought. If citation counts are sought in addition to *h* scores, then Scopus is preferable since it will identify more complete citation data. In the latter case, Web of Science can be used as an additional data source to account for pre-1996 citations, if needed.

While the selection of a database for a citation ranking study of HCI researchers has no bearing on rankings, a more complete citation count of individual HCI researchers, as found in Scopus, has significant implications on mapping the scholarly/scientific impact of these researchers. For example, looking at the results of the top three cited researchers (Rogers, Benford, and Rodden), it was found that there are significant differences between Scopus and Web of Science in terms of the identity of the top five citing authors, journals/conferences, universities, and countries. In all but three instances, the top five in Scopus varied significantly from the top five in Web of Science (see Table 7).

Regarding the *h*-index, as mentioned earlier, we generated two sets of *h* scores in each database for each researcher: one that is calculated by the database system (we called this, *system count*) and another based on citation searches of individual works (we called this, *manual count*). We also generated a system count and a manual count of *h* scores based on the union of data from both databases; this was done in order to assess the value and necessity of using multiple data sources in calculating *h* scores. Our results show that manually-based *h* counts in both Scopus and Web of Science generate significantly higher *h* scores of individual researchers than system-based *h* counts (see Table 8). This was not surprising because by definition manual *h* scores will always be equal or greater than the system count. This is so because the former takes into account all works produced or published by the researchers (in this case 1,440 journal articles, conference papers, book chapters, and so on) whereas the latter relies on only those items covered or indexed by the databases (in this case 647 or 45% of the 1,440 works produced/published by the researchers). These findings suggest that databases relied on to automatically calculate *h* scores must be used and interpreted with extreme caution (see Figure 3 and, for more detail, Table 8), particularly because the differences in the two counting methods vary significantly from one researcher to the other (from a low 50% to a high 200%). These major differences between the two counting methods imply that even when comparing researchers from the same domain, one should use the manually-based count method rather than the system-based count method for calculating *h* scores.

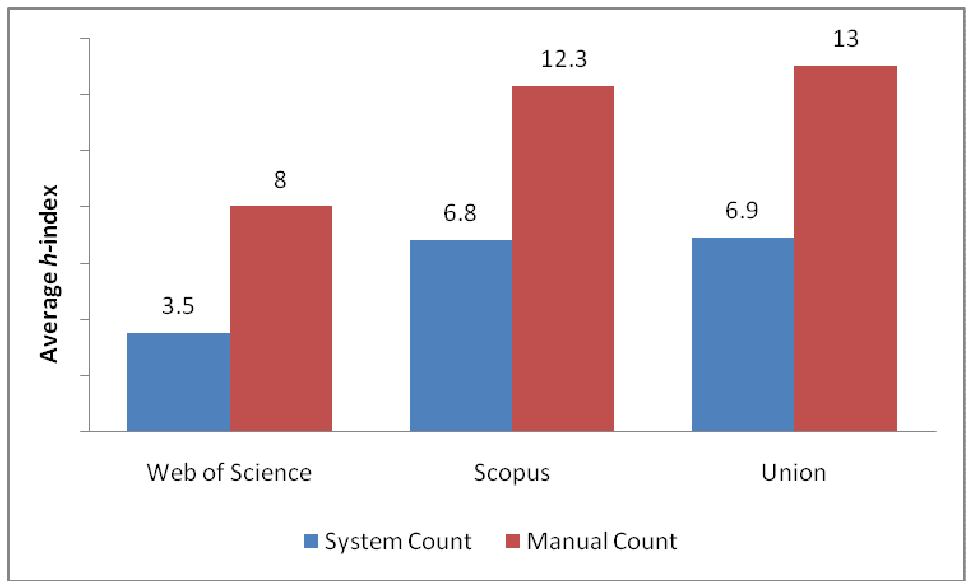

**Figure 3. Average *h* scores of study sample by counting method**

Our results additionally show that Scopus not only generates significantly higher *h* scores than Web of Science (regardless of the counting method used—system or manual), but Scopus also differentiates between the researchers in a more nuanced fashion as illustrated in the difference between top ranked and bottom ranked (variance in Web of Science equals 11 in comparison to 16 in Scopus). Results also show that the addition of citations from Web of Science to those of Scopus does not significantly alter the *h* scores or rankings of the researchers, implying that it would be unnecessary to use both databases to generate *h* scores of HCI researchers. This is an important finding particularly because it is extremely tedious and labor-intensive to generate *h* scores based on the union of citations from two databases.

In summary, our findings suggest that broader coverage of literature by citation databases does make a significant difference on citation counts, citation mapping (as illustrated with the examples provided in Table 7), and *h* scores of individual researchers in HCI. Future research should explore whether this is true in other domains.

**Comparison with Google Scholar**

Given the growing popularity of Google Scholar as a citation analysis tool (e.g., Golderman & Connolly, 2007; Neuhaus, Neuhaus, & Asher, 2008), we decided to compare the *h*-index scores derived from Google Scholar with those from the union of Scopus and Web of Science. The reasons for doing this include: (1) Google Scholar can be used to generate *h* scores for an author in a matter of seconds or minutes (especially when using such tools as Harzing's Publish or Persih—http://www.harzing.com/), in comparison to hours in the case of Scopus and Web of Science's manual counts. (2) Google Scholar's scores are based on a much larger body of literature than that of Scopus and Web of Science combined. (3) Google Scholar is a freely available tool as opposed to the very expensive, subscription-based Scopus and Web of Science, allowing many researchers with limited access to utilize and apply some citation-based exercises. And (4) Google Scholar generates manual type of *h* scores rather than system type of *h* scores. If *h*-index studies consistently find positive correlations between Google Scholar's *h* scores and those of manually calculated scores in Scopus and/or Web of Science, one could potentially use Google

Scholar as a possible alternative, especially when all things being equal (e.g., comparing authors within the same research domain).

In this study, results showed a very significant correlation between the *h*-index ranking in Google Scholar with that of the union of Scopus and Web of Science—Spearman rank order correlation coefficient for the two rankings being 0.960 (see Table 9). The main difference between the two rankings is that Google Scholar helps distinguish between the researchers in a more nuanced fashion than the union of Scopus and Web of Science, as evidenced by the larger variance between top ranked and bottom ranked researchers (30 in comparison to 18, respectively). This was not surprising because, unlike Scopus and Web of Science which cover only journal items and conference papers, Google Scholar additionally covers books, book chapters, dissertations, theses, reports, and conference workshops and presentations, among others, without any geographic or linguistic limitations. According to Meho and Yang (2007), approximately one-fourth of all Google Scholar citations in the field of library and information science come from these latter types of sources and nearly one-fourth of Google Scholar's citations are identified through full-text documents made available online by their authors (i.e., self-archived) rather than from official sources. It is these sources of citations that contribute to the large discrepancy in *h* scores between Google Scholar and the union of Scopus and Web of Science. It is also these same sources that one must pay attention to when interpreting Google Scholar-based *h* scores because their quality is not the same as the quality of journals and conferences covered by the commercial citation databases.

**CONCLUSIONS AND IMPLICATIONS**

This study shows that, in HCI, conference proceedings constitute, along with journals, a major channel of written communication. Many of these proceedings are published by ACM and IEEE and also by Springer in the form of *Lecture Notes in Computer Science/Lecture Notes in Artificial Intelligence* series. Using a sample of 22 top HCI researchers in the UK (with backgrounds in architecture, cognitive science, computer science, design, engineering, ergonomics, human factors, psychology, sociology, and software engineering), the study illustrated the necessity of using Scopus instead of Web of Science for citation-based research and evaluation in HCI. In addition to providing significantly more comprehensive coverage of relevant and high impact publications and generating more complete citation counts of individual HCI scholars, Scopus produces significantly higher $h$ scores for these scholars than Web of Science does. The addition of Web of Science citation data to those of Scopus virtually does not alter the $h$ scores of HCI researchers. The study also illustrated the necessity of manually identifying individual scholar's $h$ scores rather than relying on scores automatically calculated by the databases.

Our findings corroborate results found in other studies regarding the inappropriateness of using Web of Science exclusively as a source of bibliometrics data in domains where conference proceedings represent a major channel of written communication (e.g., computer science). Although more studies are needed to compare the citation coverage of Scopus and Web of Science in various domains, this paper highlights the significance of the selection and use of appropriate data sources and $h$-index counting methods in conducting citation-based research and evaluation. Bibliometricians, academic departments, research centers, administrators, and governments can no longer limit themselves to Web of Science because they are familiar with it, have access to it, or because it is the more established data source. Today, there are many other databases to choose from as sources of citation data. A challenge is to systematically explore these data sources and to determine which one(s) are better for what research domains. This is very important to emphasize because identifying citation counts and calculating $h$ scores using data from two or more databases can be quite labor-intensive and, in many cases, unnecessary. Still, the use of appropriate data sources and methodologies is necessary to generate valid and reliable results and make accurate or more informed research assessments.

Moreover, regardless of which citation database(s) or data source(s) are used, the principles of bibliometrics research should be observed (see Weingart, 2005): (1) it has to be applied by professional people with theoretical understanding and thorough technical knowledge of the databases, retrieval languages, and the abbreviations, concepts, and/or terminologies of the domain under investigation; (2) it should only be used in accordance with the established principles of "best practice" of professional bibliometrics as described by van Raan (1996); and (3) it should only be applied in conjunction with qualitative peer review.

The emergence of Scopus, Google Scholar, and dozens of citation-enhanced databases (see Ballard & Henry, 2006; Golderman & Connolly, 2007; Roth, 2005) will help provide better services from the producers of these databases as they compete for clients and market share. Such competition will compel database producers to pay more attention towards providing higher quality data in the form of clean and correct citations, and more complete literature coverage. As far as Web of Science is concerned, if it were to improve its literature coverage of HCI, this study recommends that it indexes those high-impact journals and conference proceedings identified in this study (see Table 5).

**Acknowledgments:** We are grateful to Blaise Cronin, Alice Robbin, and three anonymous reviewers for their comments on the paper, to Hazel Glover for her help with providing publication lists, and to Elsevier for providing us with access to Scopus. We are also grateful for the support from the Equator Interdisciplinary Research Collaboration (http://www.equator.ac.uk) which was funded by the UK's Engineering and Physical Sciences Research Council (EPSRC GR/N15986/01).

**CITED REFERENCES**

Adam, D. (2002). The counting house. *Nature, 415* (6873), 726.729.

Bakkalbasi, N., Bauer, K., Glover, J., & Wang, L. (2006). Three options for citation tracking: Google Scholar, Scopus and Web of Science. *Biomedical Digital Libraries, 7*. Retrieved November 25, 2007 from http://www.pubmedcentral.nih.gov/articlerender.fcgi?artid=1533854

Ball, R., & Tunger, D. (2006). Science indicators revisited – Science Citation Index versus SCOPUS: A bibliometric comparison of both citation databases. *Information Services & Use, 26*(4), 293–301.

Ballard, S., & Henry, M. (2006). Citation searching: New players, new tools. *The Searcher, 14*(9), 24-33.

Bar-Ilan, J. (2006). An ego-centric citation analysis of the works of Michael O. Rabin based on multiple citation indexes. *Information Processing and Management, 42*(6), 1553-1566.

Bar-Ilan, J. (2008a). Informetrics at the beginning of the 21st century—A review. *Journal of Informetrics, 2*(1), 1-52.

Bar-Ilan, J. (2008b). Which h-index? – A comparison of WoS, Scopus and Google Scholar. *Scientometrics, 74*(2), 257-271.

Bar-Ilan, J., Levene, M., & Lin, A. (2007). Some measures for comparing citation databases. *Journal of Informetrics, 1*(1), 26-34.

Bauer, K., & Bakkalbasi, N. (2005). An examination of citation counts in a new scholarly communication environment. *D-Lib Magazine, 11*(9). Retrieved November 25, 2007, from http://www.dlib.org/dlib/september05/bauer/09bauer.html

Borgman, C. L. (ed.). (1990). *Scholarly communication and bibliometrics*. Newbury Park: Sage Publications.

Borgman, C. L., & Furner, J. (2002). Scholarly communication and bibliometrics. *Annual Review of Information Science and Technology, 36*, 3-72.

Bornmann, L., Mutz, R., & Daniel, H-D. (2008). Are there better indices for evaluation purposes than the *h* index? A comparison of nine different variants of the *h* index using data from biomedicine. *Journal of the American Society for Information Science and Technology, 59*(5), 830-837.

Butler, L. (2007). Assessing university research: a plea for a balanced approach. *Science and Public Policy, 34*(8), 565-574.

Cronin, B. (1996). Rates of return to citation. *Journal of Documentation, 52*(2), 188-197.

Cronin, B., & Meho, L. I. (2008). The shifting balance of intellectual trade in information studies. *Journal of the American Society for Information Science and Technology, 59*(4), 551–564.

Cronin, B., & Meho, L. I. (2006). Using the *h*-index to rank influential information scientists. *Journal of the American Society for Information Science and Technology, 57*(9), 1275-1278.

Dillon, T. W. (1995). Mapping the discourse of HCI researchers with citation analysis. *ACM SIGCHI Bulletin, 27*(4), 56-62.

Falagas, M. E., Pitsouni, E. I., Malietzis, G. A., & Pappas, G. (2008). Comparison of PubMed, Scopus, Web of Science, and Google Scholar: Strengths and weaknesses. *The FASEB Journal, 22*(2), 338-342.

Galliers, R. D., & Whitley, E. A. (2002). An anatomy of European information systems research ECIS 1993-ECIS 2002: Some initial findings. In S. Wrycza (Ed.), Information systems and the future of the digital economy, Gdansk, Poland, 6th - 8th June 2002: ECIS 2002: proceedings of the Xth European Conference on Information Systems, pp. 3–18.

Garfield, E. (1979). *Citation indexing: Its theory and application in science, technology, and humanities.* New York: Wiley.

Golderman, G., & Connolly, B. (2007). Who cited this? Library Journal (Net Connect), 132, 18-26.

Goodman, D., & Deis, L. (2007). Update on Scopus and Web of Science. *The Charleston Advisor*, 8(3), 15-18.

Goodrum, A. A., McCain, K. W., Lawrence, S., & Giles, C. L. (2001). Scholarly publishing in the Internet age: A citation analysis of computer science literature. *Information Processing & Management, 37*(5), 661–675.

Grudin, J. (2008). A moving target: The evolution of HCI. In A. Sears & J. A. Jacko (Eds.), *The human-computer interaction handbook: Fundamentals, evolving technologies, and emerging applications*, 2nd ed. (pp. 1-24). New York: Lawrence Erlbaum Associates.

Hewett, T. T., Baecker, R., Card, S., Carey, T., Gasen, J., Mantei, M., Perlman, G., Strong, G., & Verplank, W. (1992). ACM *SIGCHI Curricula for Human-Computer Interaction*. Retrieved March 03, 2008 from http://sigchi.org/cdg/cdg2.html

Hirsch, J. E. (2005). An index to quantify an individual's scientific research output. *Proceedings of the National Academy of Sciences of the United States of America, 102*(46), 16569-16572.

Jacsó, P. (2005). As we may search—comparison of major features of the Web of Science, Scopus, and Google Scholar citation-based and citation-enhanced databases. *Current Science, 89*(9), 1537-1547. Retrieved November 25, 2007, from http://www.ias.ac.in/currsci/nov102005/1537.pdf

Jacsó, P. (2006). Deflated, inflated and phantom citation counts. *Online Information Review, 30*(3), 297-309.

Jin, B. H., Liang, L. M., & Rousseau R. (2007). The R- and AR-indices: Complementing the h-index. *Chinese Science Bulletin, 52*(6), 855-863.

Kousha, K., & Thelwall, M. (2006). Sources of Google Scholar citations outside the Science Citation Index: A comparison between four science disciplines. In *Book of Abstracts, 9th International Science & Technology Indicators Conference*, Leuven, Belgium, 72–73.

Meho, L. I. (2007). The rise and rise of citation analysis. *Physics World, 20*(1), 32-36.
24

Meho, L. I., & Yang, K. (2007). Impact of data sources on citation counts and rankings of LIS faculty: Web of Science vs. Scopus and Google Scholar. *Journal of the American Society for Information Science and Technology, 58*(13), 2105-2125.

Moed, H. F. (2005). *Citation analysis in research evaluation*. Dordrecht, The Netherlands: Springer.

Moed, H. F. (2007). The future of research evaluation rests with an intelligent combination of advanced metrics and transparent peer review. *Science and Public Policy, 34*(8), 575-583.

Moed, H. F. (2008). UK Research Assessment Exercises: Informed judgments on research quality or quantity? *Scientometrics, 74*(1), 153-161.

Moed, H. F., & Visser, M. S. (2007). *Developing bibliometric indicators of research performance in computer science: An exploratory study*. Centre for Science and Technology Studies (CWTS). Leiden University, the Netherlands. Research Report to the Council for Physical Sciences of the Netherlands Organisation for Scientific Research (NWO). CWTS Report 2007-01. Retrieved November 25, 2007, from http://www.cwts.nl/cwts/NWO_Inf_Final_Report_V_210207.pdf

Neary, J. P., Mirrlees, J. A., & Tirole, J. (2003). Evaluating economics research in Europe: An introduction. *Journal of the European Economic Association, 1*(6), 1239-1249.

Neuhaus, C., Neuhaus, E., & Asher, A. (2008). Google Scholar Goes to School: The Presence of Google Scholar on College and University Web Sites. *The Journal of Academic Librarianship, 34*(1), 39-51.

Norris, M., & Oppenheim, C. (2003). Citation counts and the Research Assessment Exercise V. *Journal of Documentation, 59*(6), 709-730.

Norris, M., & Oppenheim, C. (2007). Comparing alternatives to the Web of Science for coverage of the social sciences' literature. *Journal of Informetrics, 1*(2), 161-169.

Noruzi, A. (2005). Google Scholar: The New Generation of Citation Indexes. *Libri, 55*(4), 170-180.

Oppenheim, C. (2007). Using the *h*-Index to rank influential British researchers in information science and librarianship. *Journal of the American Society for Information Science and Technology, 58*(2), 297-301.

Oulasvirta, A. (2007). CHI statistics. Retrieved November 25, 2007, from http://www.hiit.fi/node/290

Palacios-Huerta, I., & Volij, O. (2004). The measurement of intellectual influence. *Econometrica, 72*(3), 963-977.

Pinski, G., & Narin, F. (1976). Citation Influence for Journal Aggregates of Scientific Publications: Theory, with Application to the Literature of Physics. *Information Processing & Management, 12*(5), 297-312.

Rinia, E. J., van Leeuwen, T. N., van Vuren, H. G., & van Raan, A. F. J. (1998). Comparative analysis of a set of bibliometric indicators and central peer review criteria: Evaluation of condensed matter physics in the Netherlands. *Research Policy, 27*(1), 95-107.
25

**Table 1. Academic background of the study sample**

| Name | Year Ph.D. completed | University | Country | Field/Discipline |
|---|---|---|---|---|
| Barkhuus, Louise | 2004 | IT University of Copenhagen | Denmark | Computer Science |
| Benford, Steve D.* | 1988 | University of Nottingham | United Kingdom | Computer Science |
| Brown, Barry A. T. | 1998 | University of Southampton | United Kingdom | Sociology |
| Chalmers, Matthew* | 1989 | University of East Anglia | United Kingdom | Computer Science |
| Cheverst, Keith W. J. | 1999 | University of Lancaster | United Kingdom | Computer Science |
| Crabtree, Andy | 2001 | University of Lancaster | United Kingdom | Sociology |
| De Roure, David C.* | 1990 | University of Southampton | United Kingdom | Computer Science |
| Fitzpatrick, Geraldine* | 1998 | University of Queensland | Australia | Computer Science & Electrical Engineering |
| Friday, Adrian J.* | 1996 | University of Lancaster | United Kingdom | Computer Science |
| Gaver, William W.* | 1988 | University of California at San Diego | United States | Cognitive Science |
| Gellersen, Hans W.* | 1996 | University of Karlsruhe | Germany | Software Engineering |
| Izadi, Shahram | 2004 | University of Nottingham | United Kingdom | Computer Science |
| Muller, Henk L.* | 1993 | University of Amsterdam | The Netherlands | Computer Science |
| Price, Sara | 2001 | University of Sussex | United Kingdom | Psychology |
| Randell, Cliff** | 2007 | University of Bristol | United Kingdom | Computer Science |
| Rodden, Tom A.* | 1990 | University of Lancaster | United Kingdom | Computer Science |
| Rogers, Yvonne* | 1988 | University of Wales at Swansea | United Kingdom | Psychology |
| Schmidt, Albrecht | 2002 | University of Lancaster | United Kingdom | Computer Science |
| Schnädelbach, Holger | 2007 | University College London | United Kingdom | Architecture |
| Stanton-Fraser, Danaë E. B. | 1997 | University of Leicester | United Kingdom | Psychology |
| Steed, Anthony* | 1996 | Queen Mary, University of London | United Kingdom | Computer Science |
| Weal, Mark J. | 2000 | University of Southampton | United Kingdom | Computer Science |

*Principal Investigator.
**Actively publishing since 2000.

**Table 2. Total citations by document type (1996-2007)**

| Document Type | Web of Science | | Scopus | | Union of Web of Science and Scopus | |
|---|---|---|---|---|---|---|
| | Count | % | Count | % | Count | % |
| Journal articles | 2,833 | 71% | 3,584 | 52% | 3,850 | 52% |
| Conference papers | 1,029 | 26% | 3,207 | 46% | 3,416 | 46% |
| Review articles | 72 | 2% | 76 | 1% | 86 | 1% |
| Editorial materials | 64 | 2% | 48 | 1% | 71 | 1% |
| Other | 13 | 0% | 4 | 0% | 16 | 0% |
| **Total** | **4,011** | **101%*** | **6,919** | **100%** | **7,439** | **100%** |
| **Total from Journals** | **2,982** | **74%** | **3,712** | **54%** | **4,023** | **54%** |
| **Total from Proceedings** | **1,029** | **26%** | **3,207** | **46%** | **3,416** | **46%** |

*The total percent is over 100% due to rounding.

**Table 3. Unique citations by document type (1996-2007)**

| Document Type | Unique citations in both databases | |
|---|---|---|
| | Count | % |
| Journal articles (n=4,023) | 1,352 | 34% |
| Conference papers (n=3,416) | 2,596 | 76% |
| Total (n=7,439) | 3,948 | 53% |

**Table 4. Overlap in citations by document type (1996-2007)**

| Document Type | Overlap between Web of Science and Scopus | |
|---|---|---|
| | Count | % |
| Journal articles (n=4,023) | 2,671 | 66% |
| Conference papers (n=3,416) | 820 | 24% |
| Total (n=7,439) | 3,491 | 47% |

**Table 5. Top 42 sources of citations by database (1996-2007)**

| Rank | Sources of citations | Web of Science | Scopus | Union of Web of Science and Scopus | Scopus IF (rank) | JCR Impact Factor |
|---|---|---|---|---|---|---|
| **JOURNALS** | | | | | | |
| 1 | Presence: Teleoperators and Virtual Environments | 156 | 155 | 159 | 1.480 (10) | 1.000 |
| 2 | International Journal of Human-Computer Studies | 130 | 123 | 131 | 1.615 (7) | 1.094 |
| 3 | Interacting with Computers | 113 | 105 | 115 | 1.140 (17) | 0.833 |
| 4 | **Computer Supported Cooperative Work** | | **91** | **91** | **2.000 (4)** | **NA** |
| 5T | Cyberpsychology & Behavior | 53 | 53 | 60 | 1.269 (13) | 1.061 |
| 5T | IEEE Pervasive Computing | 50 | 48 | 60 | 2.971 (2) | 2.062 |
| 5T | Personal and Ubiquitous Computing* | 48 | 57 | 60 | 1.427 (12) | NA |
| 8 | Behaviour & Information Technology | 57 | 58 | 59 | 1.097 (19) | 0.743 |
| 9 | Journal of the American Society for Information Science and Technology | 53 | 46 | 55 | 1.766 (6) | 1.555 |
| 10 | Human-Computer Interaction | 44 | 41 | 46 | 3.043 (1) | 2.391 |
| 11T | Computer Networks | 36 | 29 | 39 | 1.200 (14) | 0.631 |
| 11T | International Journal of Human-Computer Interaction | 39 | 36 | 39 | 0.695 (21) | 0.431 |
| 13T | Communications of the ACM | 33 | 30 | 35 | 1.991 (5) | 1.509 |
| 13T | Computers & Education | 34 | 31 | 35 | 1.464 (11) | 1.085 |
| 15 | Information and Software Technology | 31 | 25 | 31 | 1.138 (18) | 0.726 |
| 16 | Computers & Graphics | 27 | 25 | 28 | 0.953 (20) | 0.601 |
| 17 | Information Processing & Management | 24 | 22 | 25 | 1.576 (8) | 1.546 |
| 18 | IEEE Multimedia | 22 | 24 | 24 | 1.148 (16) | 1.317 |
| 19T | **ACM Transactions on Computer-Human Interaction** | | **23** | **23** | **2.861 (3)** | **NA** |
| 19T | IEEE Computer Graphics and Applications | 23 | 17 | 23 | 1.556 (9) | 1.429 |
| 19T | Journal of Computer Assisted Learning | 23 | 23 | 23 | 1.163 (15) | 0.532 |
| 19T | **New Review of Hypermedia and Multimedia** | | **23** | **23** | **0.565 (22)** | **NA** |
| | Total number of citations (% of all citations in journals) | 996 (33%) | 1,085 (29%) | 1,184 (29%) | | |
| **CONFERENCE PROCEEDINGS** | | | | | | |
| 1 | **ACM Conference on Human Factors in Computing Systems** | | **211** | **211** | **2.478 (1)** | |
| 2 | **ACM Conference on Computer Supported Cooperative Work** | | **72** | **72** | **-** | |
| 3 | Ubicomp: Ubiquitous Computing, Proceedings (LNCS) | 67 | 55 | 69 | - | |
| 4 | **IEEE International Conference on Pervasive Computing and Communications, PerCom** | | **64** | **64** | **0.934 (3)** | |
| 5 | **Proceedings of SPIE - The International Society for Optical Engineering** | | **60** | **60** | **-** | |
| 6 | ACM Conference on Human-Computer Interaction with Mobile Devices and Services, MobileHCI (LNCS) | 45 | 48 | 58 | - | |
| 7 | **IEEE Virtual Reality Conference** | | **51** | **51** | **0.612 (5)** | |

| Rank | Sources of citations | Web of Science | Scopus | Union of Web of Science and Scopus | Scopus IF (rank) | JCR Impact Factor |
|---|---|---|---|---|---|---|
| **8** | **ACM Conference on Hypertext and Hypermedia**\*\* | | **49** | **49** | **0.915 (4)** | |
| **9** | **ACM Conference on Designing Interactive Systems, DIS** | | **45** | **45** | **-** | |
| 10 | On The Move to Meaningful Internet Systems Conference (LNCS) | 26 | 34 | 39 | 0.155 (11) | |
| **11** | **ACM International Conference on Collaborative Virtual Environments** | | **34** | **34** | **-** | |
| 12T | Human-Computer Interaction – INTERACT (LNCS) | 18 | 33 | 33 | - | |
| 12T | International Conference on Computer Supported Cooperative Work in Design, CSCWD (LNCS) | 3 | 33 | 33 | 0.242 (9) | |
| **14T** | **ACM Symposium on Virtual Reality Software and Technology, VRST** | | **32** | **32** | **-** | |
| 14T | International Conference on Embedded and Ubiquitous Computing, EUC (LNCS) | 30 | 30 | 32 | 0.141 (12) | |
| **16** | **IEEE International Conference on Advanced Information Networking and Application, AINA** | | **30** | **30** | **0.469 (7)** | |
| 17 | IEEE International Conference on Pervasive Computing, PERVASIVE (LNCS)\* | 28 | 27 | 28 | 1.500 (2) | |
| **18** | **ACM Symposium on Applied Computing** | | **26** | **26** | **0.521 (6)** | |
| 19T | The Semantic Web: International Semantic Web Conference, ISWC (LNCS)\* | 20 | 24 | 25 | 0.323 (8) | |
| **19T** | **IEEE International Conference on Systems, Man and Cybernetics** | | **24** | **24** | **0.178 (10)** | |
| | Total number of citations (% of all citations in conference proceedings) | 237 (23%) | 982 (31%) | 1,015 (30%) | | |
| | **Grand Total (% of all citations in database)** | **1,233 (31%)** | **2,067 (30%)** | **2,219 (30%)** | | |

- The figures in the Scopus, Web of Science, and the Union of Scopus and Web of Science columns refer to the number of citations found in each journal or conference proceeding to the works of the 22 researchers.
- The figures in the Scopus IF and JCR IF columns refer to the citing sources' 2006 impact factor scores.
- NA = Not available.
- Items in bold are those citing journals and conference proceedings covered exclusively in Scopus.
- LNCS stands for the *Lecture Notes in Computer Science/ Lecture Notes in Artificial Intelligence* series, published by Springer.
- The IF scores were calculated in Scopus as follows: (Cites in 2006 to articles published in 2005 + Cites in 2006 to articles published in 2004) divided by Number of articles published in 2004-2005. Unique citations found through the "More" tab/search feature were accounted for in the IF calculations. We used Scopus instead of Thomson Scientific *Journal Citation Reports* (*JCR*) because the latter covers only 18 of the 42 sources in question in comparison to 34 in Scopus. We could not calculate the impact factor for eight conference proceedings because of coverage irregularities by Scopus and/or because some proceedings are published once every two years instead of annually. The correlation between IF scores in *JCR* and those in Scopus of the 18 titles commonly covered by both sources was found to be statistically significant with Spearman rank order correlation coefficient of 0.876.

\*2007 IF.
\*\*2004 IF.

**Table 6. Citation counts and rankings of researchers (1996-2007)**

| Name | Web of Science | | Scopus | | Difference | | Union of Web of Science and Scopus | |
|---|---|---|---|---|---|---|---|---|
| | Count | Ranking | Count | Ranking | Count (Percent) | Ranking | Count | Ranking |
| Rogers* | 753 | 1 | 1,229 | 1 | 476 (63%) | 0 | 1,319 | 1 |
| Benford* | 572 | 3 | 1,179 | 2 | 607 (106%) | 1 | 1,244 | 2 |
| Rodden* | 577 | 2 | 1,075 | 3 | 498 (86%) | -1 | 1,138 | 3 |
| De Roure* | 421 | 5 | 764 | 4 | 343 (81%) | 1 | 834 | 4 |
| Gaver* | 427 | 4 | 704 | 5 | 277 (65%) | -1 | 753 | 5 |
| Friday* | 348 | 8 | 649 | 6 | 301 (86%) | 2 | 677 | 6 |
| Schmidt | 329 | 9 | 607 | 7 | 278 (84%) | 2 | 654 | 7 |
| Gellersen* | 311 | 10 | 591 | 8 | 280 (90%) | 2 | 627 | 8 |
| Cheverst | 352 | 7 | 586 | 9 | 234 (66%) | -2 | 618 | 9 |
| Steed* | 354 | 6 | 584 | 10 | 230 (65%) | -4 | 615 | 10 |
| Chalmers* | 256 | 11 | 414 | 11 | 158 (62%) | 0 | 442 | 11 |
| Crabtree | 136 | 14 | 326 | 12 | 190 (140%) | 2 | 334 | 12 |
| Stanton-Fraser | 197 | 12 | 305 | 14 | 108 (55%) | -2 | 320 | 13 |
| Brown | 155 | 13 | 308 | 13 | 153 (99%) | 0 | 318 | 14 |
| Fitzpatrick* | 98 | 17 | 209 | 15 | 111 (113%) | 2 | 227 | 15 |
| Muller* | 113 | 15 | 199 | 17 | 86 (76%) | -1 | 212 | 16T |
| Weal | 102 | 16 | 203 | 16 | 101 (99%) | -1 | 212 | 16T |
| Randell | 81 | 18 | 171 | 18 | 90 (111%) | 0 | 179 | 18 |
| Izadi | 68 | 19 | 160 | 19 | 92 (135%) | 0 | 168 | 19 |
| Barkhuus | 60 | 20 | 125 | 20 | 65 (108%) | 0 | 130 | 20 |
| Schnädelbach | 38 | 21 | 85 | 21 | 47 (124%) | 0 | 87 | 21 |
| Price | 31 | 22 | 68 | 22 | 37 (119%) | 0 | 69 | 22 |
| **TOTAL (excluding overlap)** | **4,011** | | **6,919** | | **2,908 (73%)** | | **7,439** | |

*Principal Investigator.

**Table 7. Differences between Scopus and Web of Science in terms of top citing entities of the three most cited researchers**

| Researcher | Web of Science | Scopus | % Mismatch |
|---|---|---|---|
| | **Top Citing Authors** | | |
| **Benford** | Pilar Herrero (10)<br>Chris Greenhalgh (6)<br>Ling Chen (5)<br>Jin Zhang (5)<br>Paul Luff (4)<br>Minh Hong Tran (4) | Pilar Herrero (13)<br>Ling Chen (10)<br>Andy Crabtree (10)<br>Azzedine Boukerche (8)<br>Carl Gutwin (7) | 64% |
| **Rodden** | Steve Benford (6)<br>John M. Carroll (5)<br>Yvonne Rogers (5)<br>Jeremy N. Bailenson (4)<br>Paul Dourish (4)<br>Yan Huang (4)<br>Paul F. Marty (4) | Andy Crabtree (14)<br>Steve Benford (10)<br>Paul Dourish (8)<br>David Martin (7)<br>Jeremy N. Bailenson (5)<br>Barry Brown (5)<br>Alan Dix (5)<br>Rahat Iqbal (5)<br>Marianne Petersen (5)<br>Yvonne Rogers (5)<br>Michael B. Twidale (5) | 56% |
| **Rogers** | Andrew Large (6)<br>Peter C. -H. Cheng (4)<br>Marian Petre (4)<br>Yin-Leng Theng (4)<br>Daniella Petrelli (4)<br>Ping Zhang (4) | Andrew Large (7)<br>Gloria Mark (6)<br>Mark J. Weal (6)<br>Paloma Diaz (5)<br>John D. Fernandez (5)<br>Athanasis Karoulis (5)<br>Toni Robertson (5) | 85% |
| | **Top Citing Sources** | | |
| **Benford** | Presence: Teleoperators and Virtual Environments (57)<br>UbiComp (21)<br>International Journal of Human-Computer Studies (15)<br>Interacting with Computers (14)<br>Personal and Ubiquitous Computing (11) | CHI Conference (58)<br>Presence: Teleoperators and Virtual Environments (57)<br>Int. Conf. on Collaborative Virtual Environments (32)<br>Computer Supported Cooperative Work (31)<br>IEEE Virtual Reality Conference (22) | 80% |
| **Rodden** | Presence: Teleoperators and Virtual Environments (26)<br>International Journal of Human-Computer Studies (24)<br>UbiComp (23)<br>Interacting with Computers (20)<br>Personal and Ubiquitous Computing (17) | Computer Supported Cooperative Work (59)<br>CHI Conference (54)<br>ACM Conf. on Computer Supported Cooperative Work (33)<br>Presence: Teleoperators and Virtual Environments (27)<br>International Journal of Human-Computer Studies (23) | 60% |
| **Rogers** | International Journal of Human-Computer Studies (51)<br>Interacting with Computers (51)<br>Behaviour & Information Technology (26)<br>JASIST (21)<br>Computers & Education (18) | International Journal of Human-Computer Studies (49)<br>Interacting with Computers (48)<br>CHI Conference (30)<br>Behaviour & Information Technology (27)<br>JASIST (19) | 20% |
| | **Top Citing Institutions*** | | |
| **Benford** | University of Nottingham (33)<br>University of Sussex (14)<br>Lancaster University (11)<br>Universidad Politécnica de Madrid (10)<br>King's College London (8) | University of Nottingham (80)<br>University of Ottawa (23)<br>University College London (21)<br>Zhejiang University (19)<br>Fraunhofer-Gesellschaft (16)<br>Georgia Institute of Technology (16)<br>Lancaster University (16) | 67% |
| **Rodden** | Lancaster University (31)<br>University of Nottingham (21)<br>Fraunhofer-Gesellschaft (9)<br>Intel Corporation (8)<br>University of Illinois at Urbana-Champaign (7) | University of Nottingham (47)<br>Lancaster University (45)<br>Georgia Institute of Technology (19)<br>University of Aarhus (14)<br>University of California at Irvine (14) | 60% |
| **Rogers** | University of Sussex (14)<br>Loughborough University (13)<br>University of Nottingham (13)<br>McGill University (12)<br>Open University (12) | Indiana University Bloomington (20)<br>Open University (19)<br>University of Sussex (19)<br>University of Nottingham (14)<br>Loughborough University (13)<br>McGill University (13) | 9% |
| | **Top Citing Countries** | | |
| **Benford** | United Kingdom (158)<br>United States (127)<br>Germany (30)<br>Japan (28)<br>Australia (25) | United Kingdom (312)<br>United States (234)<br>China (69)<br>Japan (65)<br>Canada (52) | 40% |
| **Rodden** | United Kingdom (158)<br>United States (117)<br>Germany (36)<br>Italy (25)<br>Australia (23) | United Kingdom (286)<br>United States (235)<br>Germany (53)<br>Sweden (44)<br>Canada (43) | 40% |
| **Rogers** | United Kingdom (240)<br>United States (171)<br>Canada (37)<br>Scotland (35)<br>Australia (34) | United Kingdom (354)<br>United States (283)<br>Canada (61)<br>Australia (60)<br>Germany (48) | 20% |

The figures in parentheses refer to number of citations.
* The percentage of mismatch would have been even higher had we removed citations from the home institution of the researchers.

**Table 8. (2008) *h*-index scores of researchers (entire career)**

|  | Web of Science | | Scopus | | Union of Web of Science and Scopus | | Percent of difference between system and manual count of the union data |
| --- | --- | --- | --- | --- | --- | --- | --- |
|  | System count | Manual count | System count | Manual count | System count | Manual count |  |
| Benford* | 7 | 14 | 12 | 22 | 12 | 24 | 100% |
| Rodden* | 5 | 13 | 12 | 19 | 12 | 21 | 75% |
| Gaver* | 3 | 14 | 8 | 20 | 8 | 20 | 150% |
| De Roure* | 6 | 12 | 8 | 17 | 9 | 19 | 111% |
| Rogers* | 7 | 11 | 9 | 15 | 9 | 17 | 89% |
| Steed* | 6 | 11 | 10 | 16 | 10 | 16 | 60% |
| Gellersen* | 6 | 8 | 10 | 14 | 10 | 15 | 50% |
| Schmidt | 5 | 9 | 9 | 14 | 9 | 15 | 67% |
| Chalmers* | 2 | 7 | 8 | 13 | 8 | 13 | 63% |
| Cheverst | 5 | 9 | 7 | 12 | 7 | 13 | 86% |
| Crabtree | 2 | 7 | 8 | 13 | 8 | 13 | 63% |
| Friday* | 4 | 9 | 7 | 13 | 7 | 13 | 86% |
| Stanton-Fraser | 5 | 8 | 7 | 10 | 7 | 11 | 57% |
| Brown | 4 | 6 | 6 | 10 | 6 | 10 | 67% |
| Fitzpatrick* | 1 | 5 | 5 | 9 | 5 | 10 | 100% |
| Weal | 2 | 6 | 5 | 9 | 5 | 10 | 100% |
| Muller* | 2 | 6 | 3 | 9 | 3 | 9 | 200% |
| Randell | 1 | 5 | 4 | 9 | 4 | 9 | 125% |
| Izadi | 1 | 5 | 4 | 8 | 4 | 8 | 100% |
| Schnädelbach | 0 | 4 | 4 | 6 | 4 | 7 | 75% |
| Barkhuus | 1 | 5 | 2 | 6 | 2 | 6 | 200% |
| Price | 2 | 3 | 2 | 6 | 2 | 6 | 200% |
| **AVERAGE** | **3.5** | **8.0** | **6.8** | **12.3** | **6.9** | **13.0** | **89%** |

*Principal Investigator.

**Table 9. Comparison of *h*-index scores and rankings between Scopus and Web of Science and Google Scholar (entire career)**

| Researcher | Union of Web of Science and Scopus | | Google Scholar | | Percent of difference in scores |
|---|---|---|---|---|---|
| | **Score** | **Rank** | **Score** | **Rank** | |
| Benford* | 24 | 1 | 38 | 1T | 58% |
| Rodden* | 21 | 2 | 38 | 1T | 81% |
| Gaver* | 20 | 3 | 32 | 3 | 60% |
| De Roure* | 19 | 4 | 27 | 4T | 42% |
| Rogers* | 17 | 5 | 27 | 4T | 59% |
| Cheverst | 13 | 9T | 25 | 6T | 92% |
| Gellersen* | 15 | 7T | 25 | 6T | 67% |
| Steed* | 16 | 6 | 25 | 6T | 56% |
| Schmidt | 15 | 7T | 24 | 9 | 60% |
| Friday* | 13 | 9T | 23 | 10 | 77% |
| Chalmers* | 13 | 9T | 21 | 11 | 62% |
| Crabtree | 13 | 9T | 20 | 12 | 54% |
| Brown | 10 | 14T | 18 | 13 | 80% |
| Fitzpatrick* | 10 | 14T | 17 | 14 | 70% |
| Muller* | 9 | 17T | 15 | 15T | 67% |
| Stanton-Fraser | 11 | 13 | 15 | 15T | 36% |
| Weal | 10 | 14T | 14 | 17 | 40% |
| Randell | 9 | 17T | 13 | 18 | 44% |
| Izadi | 8 | 19 | 12 | 19 | 50% |
| Schnädelbach | 7 | 20 | 9 | 20 | 29% |
| Barkhuus | 6 | 21T | 8 | 21T | 33% |
| Price | 6 | 21T | 8 | 21T | 33% |
| **AVERAGE** | **13.0** | | **20.6** | | **59%** |